\documentclass{andp2012}
\usepackage[english]{babel}
\usepackage{color}

\keywords{Geometric quantum computation, superconducting circuits, optimal control}

\title{ Fast holonomic quantum computation on superconducting circuits with optimal control}

\author[S. Li]{Sai Li}
\author[T. Chen]{Tao Chen}
\author[Z.-Y. Xue]{Zheng-Yuan Xue \footnote{Corresponding author. E-mail:~\textsf{zyxue83@163.com}}}
\address{S. Li, T. Chen, Prof. Z.-Y. Xue\\
Guangdong Provincial Key Laboratory of Quantum Engineering and Quantum Materials,\\
GPETR Center for Quantum Precision Measurement,\\
School of Physics and Telecommunication Engineering,\\
South China Normal University,\\
Guangzhou 510006, China}
\shortauthors{S. Li, T. Chen, and Z.-Y. Xue}

\begin{abstract}
 Geometric phases induced in quantum evolutions have built-in noise-resilient characters, and thus can find applications in many robust quantum manipulation tasks.  Here, we propose a feasible and fast scheme for universal quantum computation on superconducting circuits with nonadiabatic non-Abelian geometric phases, using resonant interaction of three-level quantum system. In our scheme, arbitrary single-qubit quantum gates can be implemented in a single-loop scenario by shaping both the amplitudes and phases of the two driving microwave fields resonantly coupled to a transmon device. Moreover, nontrivial two-qubit gates can also be realized with an auxiliary transmon simultaneously coupled to the two target transmons in an effective resonant way. In particular, our proposal can be compatible to various  optimal control techniques, which further enhances the robustness of the  quantum operations. Therefore, our proposal  represents a promising way towards fault-tolerant quantum computation on solid-state quantum circuits.
\end{abstract}
\shortabstract
\begin{document}
\maketitle

\section{Introduction}
High-fidelity quantum gates are essential for the physical realization of quantum computation (QC), and thus constructing noise-resistant gate is one of the key ingredients. Meanwhile, geometric phases \cite{berry, b3,b2}, determined  by the global properties of the evolution paths, possess a kind of built-in noise-resilience feature against certain types of local noises \cite{ps1,zhu05,ps2,mj}.  Therefore, in large scale quantum systems, where control lines/devices will inevitably cause local noises, it is more promising to realize quantum manipulations in a geometric strategy \cite{gqc}. Furthermore, the non-Abelian geometric phases \cite{b3} can naturally be used to construct universal set of single-qubit gates, and together with a nontrivial two-qubit entangling gate,  holonomic QC \cite{zanardi} can be achieved.

Due to the limited coherent times of quantum systems \cite{xbwang,zhu},  the physical realization of holonomic QC based on fast nonadiabatic evolution, i.e., the nonadiabatic holonomic QC (NHQC) \cite{Sjoqvist2012, Xu2012}, is highly desirable. Recently,  the NHQC have achieved  significant theoretical \cite{vam2014, Zhang2014d, Xu2014, Xu2015, Xue2015b, es2016, Xue2016, Zhao2016, Herterich2016,xu2017, Xue2017, xu20172, zhao2017, vam2017, zhouj2018,Hong18, xu2018, vam2018,Zhang2019,nr2019} and experimental progress \cite{Abdumalikov2013,Feng2013,Zu2014,Arroyo-Camejo2014,nv2017,nv20172,li2017,Xu18,ni2018,kn2018, dje2019,yan2019,zhu2019}. Among these schemes, due to the easy  integrability and flexibility, superconducting circuits system \cite{sq2,sq3,sq4,transmon,multilevel} is recognized as a promising candidate to implement scalable QC. However, up to now, a practical implementation of NHQC on superconducting circuits is lacking, e.g., experimental demonstrations \cite{Abdumalikov2013, Xu18, yan2019} are limited to the single-qubit gates cases and the more difficult two-qubit gate do not have a feasible scheme yet, due to the need of a nonlinear quantum bus for the two target  qubits \cite{ Hong18, dje2019}. Meanwhile, this type of NHQC implementation is sensitive to the systematic error \cite{Zheng16,Jing17,Liu18}, which thus smear the main advantage of quantum computation with geometric phases.

Here, to remove the two main obstacles, we propose a practical and fast scheme to construct universal holonomic quantum gates for NHQC  on  superconducting circuits, by generalizing the previous used single time-dependent variable of Hamiltonian into a two time-dependent variables case. In our scheme, each transmon device serves as a qubit and we only use resonant sequential transitions,  driven by two microwave fields,   of the ladder type three levels in a transmon device. Meanwhile, the evolution state is set by inverse engineering of the system Hamiltonian, and thus during the cyclic evolution, it can always fulfill the time-dependent Schr\"{o}dinger equation of the govern Hamiltonian, i.e., no nonadiabatic  transitions  will occur \cite{Daems13}. In addition, at the end of the evolution, the pure geometric phase can be acquired after canceling the dynamical phase, thus the proposed gates are of the geometric nature. In this way, arbitrary single-qubit quantum gates can be implemented in a single-loop scenario by shaping both the amplitudes and phases of two microwave fields. Moreover, without introducing a quantum bus, nontrivial two-qubit gates can be realized with an auxiliary transmon simultaneously coupled to the two target transmons in an effective resonant way and the intrinsic nonlinearity  of the auxiliary transmon is large enough for our purpose. In particular, due to the inverse Hamiltonian engineering, the evolution state of our proposal has two independent variables, thus it can be compatible to various  optimal control techniques \cite{Daems13, Chen12} and further enhances the robustness of the  quantum operations.

\section{Universal single-qubit gates}
We now proceed to present our scheme based on superconducting circuits via inverse engineering Hamiltonian by solving the Schr\"{o}dinger equation  \cite{Daems13}.  Firstly, arbitrary single-qubit gates can be implemented  by using two resonant driving microwave fields with time dependent amplitudes and phases. Then, we show that our scheme is compatible with optimal control methods, thus the robustness of our implementation can be further improved.

\subsection{Inverse engineering of Hamiltonian}

In this section, we introduce how to inversely engineer the Hamiltonian based  on the Schr\"{o}dinger equation  \cite{Daems13} on superconducting circuits. In a superconducting transmon device, we consider the three lowest levels $|g\rangle$, $|e\rangle$ and $|f\rangle$, with $|g\rangle$ , $|f\rangle$ being our qubit states  and  $|e\rangle$ being an auxiliary state, to construct geometric manipulation of the device.  As shown in Fig. 1(a), two microwave fields $\Omega_{j}(t)\cos(\omega_{j}t+\phi_{j}(t))(j=1, 2)$, with  $\Omega_{j}(t)$, $\omega_{j}$ and $\phi_{j}(t)$ being the amplitudes, frequencies and phases, resonantly coupled to the sequential transitions of the three lowest levels of a transmon  \cite{Hong18}. Ignoring the higher order oscillating terms, assuming $\hbar=1$ hereafter, the effective interaction Hamiltonian can be written as
\begin{eqnarray}\label{Hd}
H_1 &=&\left[\frac{\Omega _1(t)}{2}e^{i\phi_1(t)}|g\rangle
+\frac{\Omega _2(t)}{\sqrt {2}}e^{-i\phi_2(t)}|f\rangle\right] \langle e|+\mathrm{H.c.}\notag\\
&=&\Omega e^{i\phi_1(t)} |b\rangle\langle e| + \mathrm{H.c.},
\end{eqnarray}
where $\Omega=\sqrt{(\frac{\Omega _1(t)}{2})^2+(\frac{\Omega _2(t)}{\sqrt {2}})^2}$, bright state $|b\rangle=\sin(\theta/2)|g\rangle -\cos(\theta/2)e^{-i\phi}|f\rangle$ with $\tan(\theta/2)=\frac{\Omega_1}{\sqrt{2}\Omega_2}$ and $\phi=\phi_2(t)+\phi_1(t)+\pi$ being time independent,  and dark state $|d\rangle=-\cos(\theta/2)e^{i\phi}|g\rangle-\sin(\theta/2)|f\rangle$ is decoupled.

\begin{figure}[tbp]
	\begin{center}\label{Figure1}
		\includegraphics[width=\columnwidth]{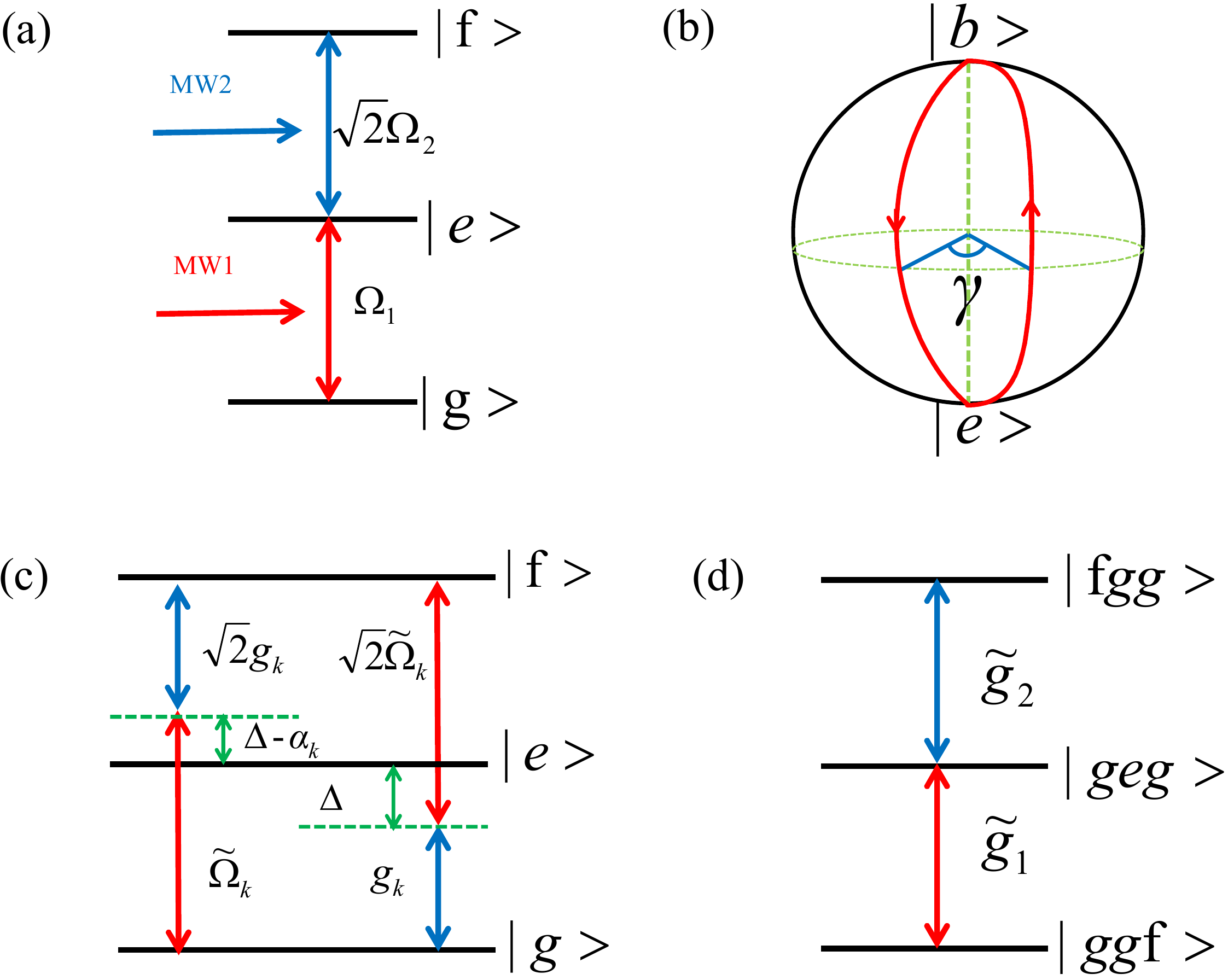}
		\caption{Illustration of the proposal. (a) The three-level configuration of the resonant situation for the single-qubit gates, with two microwave fields resonantly coupled to the three levels of a transmon device. (b) Geometric diagram of the proposed single-qubit gate. (c) Effective resonant qubit-qubit coupling configuration induced by two driving  qubits coupled to an auxiliary transmon for non-trivial two qubit gates. (d) The effective coupling configuration for two-qubit gate in the single-excitation
subspace.}
	\end{center}
\end{figure}

The Hamiltonian $ H_1$ satisfies the Schr\"{o}dinger equation of
  \begin{equation}\label{SE}
\begin{aligned}
 i\frac{\partial}{\partial t}|\varPsi(t)\rangle&=H_{1}(t)|\varPsi(t)\rangle,
\end{aligned}
\end{equation}
where the evolution state $|\varPsi(t)\rangle$ can generally be parameterized in the $\{|b\rangle, |e\rangle\}$ subspace by using two angles $\chi, \varphi$ and with a global phase $f$ as
  \begin{equation}\label{solution}
\begin{aligned}
|\varPsi(t)\rangle &=e^{-if/2}\left(\begin{array}{cc}
\cos\frac{\chi}{2} e^{-i\varphi/2}   \\
\sin\frac{\chi}{2} e^{i\varphi/2} \\
\end{array}\right).
\end{aligned}
\end{equation}
Inserting $|\varPsi(t)\rangle $ into the Schr\"{o}dinger equation, we get
\begin{equation}\label{relation}
\begin{aligned}
\dot{f}&=-\dot {\varphi}/\cos\chi,\quad\dot{\chi}=-2\Omega \sin{(\phi_1+\varphi)},\\
\dot{\varphi} &= -2 \Omega \cot{\chi}\cos{(\phi_1+\varphi)}.
\end{aligned}
\end{equation}
That is to say, if the parameters of the two microwave fields satisfy the above relations, the evolution path will just go along with the evolution state $|\varPsi(t)\rangle$, i.e., no  transitions from the  state $|\varPsi(t)\rangle$ to its orthogonal states will occur during the quantum evolution governed by Hamiltonian $H_{1}(t)$ \cite{Daems13}.

\subsection{Gates implementation and performance}
Now, we proceed to implement arbitrary holonomic single-qubit gates. A set of proper parameters $\chi$ and $\varphi$ can be chosen to realize a certain evolution, and once the parameters $\chi$ and $\varphi$ are set, the corresponding $\Omega$ and $\phi_1$ can be obtained by solving the Eq. (\ref{relation}), i.e.,
 \begin{equation}\label{Jp}
\phi_1 = \arctan{\left(\frac{\dot{\chi}}{\dot{\varphi}}\cot{\chi}\right)}-\varphi,\quad
\Omega = -\frac{\dot{\chi}}{2\sin{(\phi_1+\varphi)}},
\end{equation}
and thus fix the Hamiltonian $H_{1}(t)$. Here, we consider a cyclic evolution,  which can be achieved by setting $\chi(0)=\chi(\tau) = 0$ in Eq. (\ref{solution}), i.e., the state will start from $|b\rangle$ state and go back to it after an periodical evolution with time $\tau$, only acquiring a phase factor. During the process, the phase factor may consist of both the geometric and the the dynamical ones, and the dynamical phase is calculated to be
 \begin{equation}\label{gd}
\begin{aligned}
\gamma_d(\tau) =-\int^{\tau}_0 \langle \varPsi(t)|H_1 |\varPsi(t)\rangle dt
=\int^{\tau}_0  \frac{\dot{\varphi }\sin^2{\chi}}{2\cos{\chi}} dt.
\end{aligned}
\end{equation}
In order to induce a pure geometric phase, the dynamical phase should be zero at the end of the cyclic evolution, i.e., $\gamma_d(\tau)=0$,  which is mainly different than the previous NHQC schemes, for instance, $\gamma_d(t)=0$ should be fulfilled at any moment. However, under this strict condition, there is just one adjustable variable in the Hamiltonian which results in the previous NHQC schemes being sensitive to the systematic error \cite{Zheng16,Jing17,Liu18}. Here, our scheme make the difference in order to release the strict condition to acquire one more adjustable variable to realize more robust quantum evolution process.

In the following, we adopt a single-loop evolution path  \cite{Herterich2016,Hong18}, as illustrated in Fig. 1(b),  to induce a pure geometric phase that can be used to achieve universal single-qubit gates. Specifically, the evolution path is divided into two equal parts, in the first path $t\in [0,\tau/2]$, we set
\begin{equation}\label{parameter1}
\begin{aligned}
\chi_1 = \pi \sin^2\left({\pi t \over \tau}\right),  \quad
\varphi_1 =-\frac{\pi}{5}\sin \left ({2\pi t \over \tau} \right)-\frac{\pi}{2},
\end{aligned}
\end{equation}
the corresponding evolution operator is
$U_1 = |d\rangle \langle d| + e^{i\gamma_1}|e\rangle \langle b|+ e^{-i\gamma_1}|b\rangle \langle e|$.
In the second path $t\in [\tau/2,\tau]$, we choose
\begin{equation}\label{parameter2}
\begin{aligned}
\chi_2 = \pi \sin^2\left({\pi t \over \tau}\right),  \quad
\varphi_2 = \frac{\pi}{5}\sin \left ({2\pi t \over \tau} \right)- \gamma -\frac{\pi}{2},
\end{aligned}
\end{equation}
where $\gamma$ is an arbitrary constant angle. The corresponding evolution operator is
$U_2 = |d\rangle \langle d| + e^{i\gamma_2}|b\rangle \langle e|+ e^{-i\gamma_2}|e\rangle \langle b|$. Then, the final evolution operator can be obtained as
\begin{equation}\label{UU}
\begin{aligned}
U(\tau) =U_2U_1=|d\rangle \langle d| + e^{i\gamma}|b\rangle \langle b|
=e^{i{\frac{ \gamma}{2}}}e^{-i{\frac{ \gamma}{2}}\textbf{n}\cdot\sigma}
\end{aligned}
\end{equation}
where$\gamma = \gamma_1+\gamma_2$, $\textbf{n}=(\sin\theta\cos\phi,-\sin\theta\sin\phi,\cos\theta)$, $\sigma$ are Pauli matrices. The evolution operator is a rotation operator around the axis $\textbf{n}$ by an angle $\gamma$, can be used  to generate arbitrary single-qubit gates in a holonomic way.   Especially, we note that when $\dot{\varphi}$ or $\dot{f}$  is zero, i.e., $\chi=\pi \sin^2(\pi t/\tau)$ but $\varphi_{1,2}$ are constant, our scheme reduce to the previous  NHQC one.

\begin{figure}[tb]
	\begin{center}\label{Figure2}
		\includegraphics[width=\columnwidth]{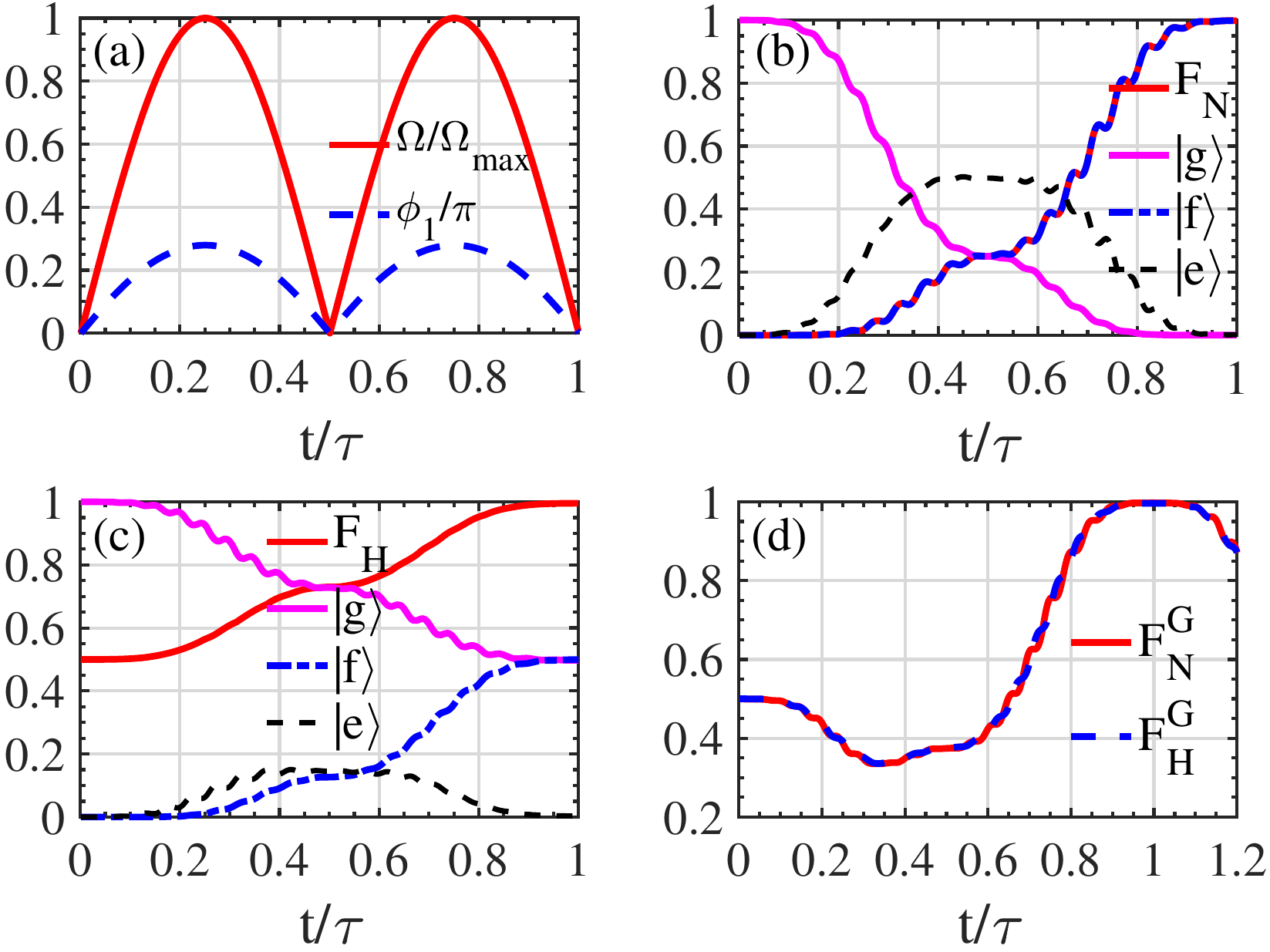}
\caption{Engineering the single-qubit gates  and their performance. (a) The parameter shapes of the Hamiltonian $H_1(t)$. (b) and (c) describe the state population and fidelity dynamics of the NOT gate  and the Hadamard gate, respectively. (d) The gate fidelity dynamics of NOT gate and Hadamard gate.}
	\end{center}
\end{figure}

The performance of the single-qubit gate can be evaluated by the Lindblad master equation of
\begin{eqnarray}  \label{master}
\dot\rho_1 &=& i[\rho_1, H_1]  + \frac{1}{2}  \left[ \Gamma_1 \mathcal{L}(\sigma_1) + \Gamma_2 \mathcal{L}(\sigma_2)\right],
\end{eqnarray}
where $\rho_1$  is the density matrix of the considered system and $\mathcal{L}(\mathcal{A})=2\mathcal{A}\rho_1 \mathcal{A}^\dagger-\mathcal{A}^\dagger \mathcal{A} \rho_1 -\rho_1 \mathcal{A}^\dagger \mathcal{A}$ is the Lindbladian of the operator $\mathcal{A}$, $\sigma_1=|g\rangle\langle e|+\sqrt{2}|e\rangle\langle f|+\sqrt{3}|f\rangle\langle h|$, $\sigma_2=|e\rangle\langle e|+2|f\rangle\langle f|+3|h\rangle\langle h|$ with $|h\rangle$ being the third excited state of a transmon, $\Gamma_1$ and $\Gamma_2$ are the corresponding decay and dephasing rates. Here, we consider the case $\Gamma_1=\Gamma_2=\Gamma= 2\pi \times 5$ kHz \cite{Barends14},  corresponding to a coherent time of 32 $\mu$s, which is well accessible with current technologies. The anharmonicity  of the transmon is set to be $\alpha=\omega_{ge}-\omega_{ef}=2\pi\times 400$  MHz \cite{jc}.
Assuming the initial state $|\psi_1\rangle=|g\rangle$ and $\tau \simeq$ 51 ns, the shapes of $\Omega$ and $\phi_1$ are shown in Fig. 2(a)  with  $\Omega_{max} = 2\pi\times 16$ MHz. Note that there will be a  bounded maximum amplitude due to the limited anharmonicity of the transmon device. Then, the NOT gate with $\theta=\pi/2$, $\phi=0$, $\gamma=\pi$ and the Hadamard gate with $\theta=\pi/4$, $\phi=0$, $\gamma=\pi$ are evaluated, using the state fidelity  defined by $F_{N/H}=_{N/H}\langle\psi_f|\rho_1|\psi_f\rangle_{N/H}$ with  $|\psi_f\rangle_N=|f\rangle$ and $|\psi_f\rangle_H=(|g\rangle+|f\rangle)/\sqrt{2}$ being the corresponding target state.  The obtained fidelities are as high as $F _{N}= 99.79\%$ and $F _{H}= 99.55\%$, as shown in  Fig. 2(b) and Fig. 2(c), respectively. The infidelity is mainly due to both the leakage caused by the small anharmonicity and relaxation and dephasing of the qubits and auxiliary state $|e\rangle$.  In addition, for a general initial state $|\psi_1\rangle=\cos\theta^{'}|g\rangle+\sin\theta^{'}|f\rangle$, the NOT and Hadamard gates
should result in an ideal final state $|\psi_f\rangle_{N}=\sin\theta^{'}|g\rangle+\cos\theta^{'}|f\rangle$ and $|\psi_f\rangle_{H}=\frac{1}{\sqrt{2}}[(\cos\theta^{'}+\sin\theta^{'})
|g\rangle+(\cos\theta^{'}-\sin\theta^{'})|f\rangle]$. To fully evaluate the performance of the implemented gates, we define the gate fidelity as $F^G_{N/H}=(\frac{1}{2\pi}){\int^{2\pi}_0}_{N/H}\langle\psi_f|\rho_1|\psi_f\rangle_{N/H} d\theta^{'}$ \cite{Poyatos97} with the integration numerically performed for 1001 input states with $\theta^{'}$ being uniformly distributed over $[0,2\pi]$. In Fig. 2(d), we have plotted the gate fidelities, where we find that the gate fidelities are $F^G _{N}= 99.75\%$ and $F^G _{H}= 99.62\%$.

\subsection{Optimal control}
Due to the parametric constrain, the previous NHQC implementations are sensitive to the systematic error \cite{Zheng16,Jing17}. Meanwhile, it is difficult to incorporate optimal control technique without additional adjustable parameters. Here, as our scheme introduce additional time dependent phase factors, we can adopt `zero systematic-error
sensitivity'-optimal protocol \cite{Chen12} to further suppress the sensitive of our implementation to the systematic error. To begin with, we consider the static systematic error situation, i.e. $\Omega\rightarrow(1+\epsilon)\Omega$. Therefore, the Hamiltonian can be written as
 \begin{equation}\label{HE}
\begin{aligned}
H_{\epsilon}(t) &=(1+\epsilon)\Omega e^{i\phi_1} |b\rangle\langle e| + \mathrm{H.c.},
\end{aligned}
\end{equation}

In our implementation, at the end of the first interval $\tau/2$, to evaluate the influence of the static systematic error,  the excitation profile is given as
 \begin{equation}\label{P}
\begin{aligned}
 P=\left| \langle \varPsi(\tau/2)|\varPsi_\epsilon(\tau/2)\rangle\right|^2
 = 1 + \tilde{O}_1+\tilde{O}_2+...,
\end{aligned}
\end{equation}
where $ |\varPsi_\epsilon(\tau/2)\rangle $ is the state with the static systematic error, and $\tilde{O}_m$ is the perturbation term of order $m$. Here, we only consider  the excitation profile $P$ to the second order, i.e.,
$P_2 = 1-\epsilon^2 q_s$,
where
\begin{equation}\label{Qs}
q_s=-\frac{\partial P_2}{\partial(\epsilon^2)}
= \left|\int^{\frac{\tau}{2}}_0 e^{-if}\dot{\chi}\sin^2{\chi} dt \right|^2
\end{equation}
represents the systematic error sensitivity. To nullify the $q_s$, we set $f(\chi) =n[2\chi - \sin{(2\chi)]}$,  $\varphi_1(0)=0$ and $\varphi_2(\tau/2)=-\gamma$, which lead to $q_s=\sin^2n\pi/(2n)^2$, i.e., for positive integer $n$, $q_s=0$. When $n\rightarrow0$,  $q_s\rightarrow \pi^2/4$,  the current implementation reduces to the previous NHQC case.  In the following numerical simulations, all the maximum value of $\Omega$ are set to be $\Omega_{max}= 2\pi\times 16$  MHz as a restriction. That is the maximum value of the optimized pulse is bounded by $\Omega_{max}$, and thus the improvement of the gate performance can only be attributed to the optimal control.
\begin{figure}[tbp]
	\begin{center}		\label{Figure3}
		\includegraphics[width=\columnwidth]{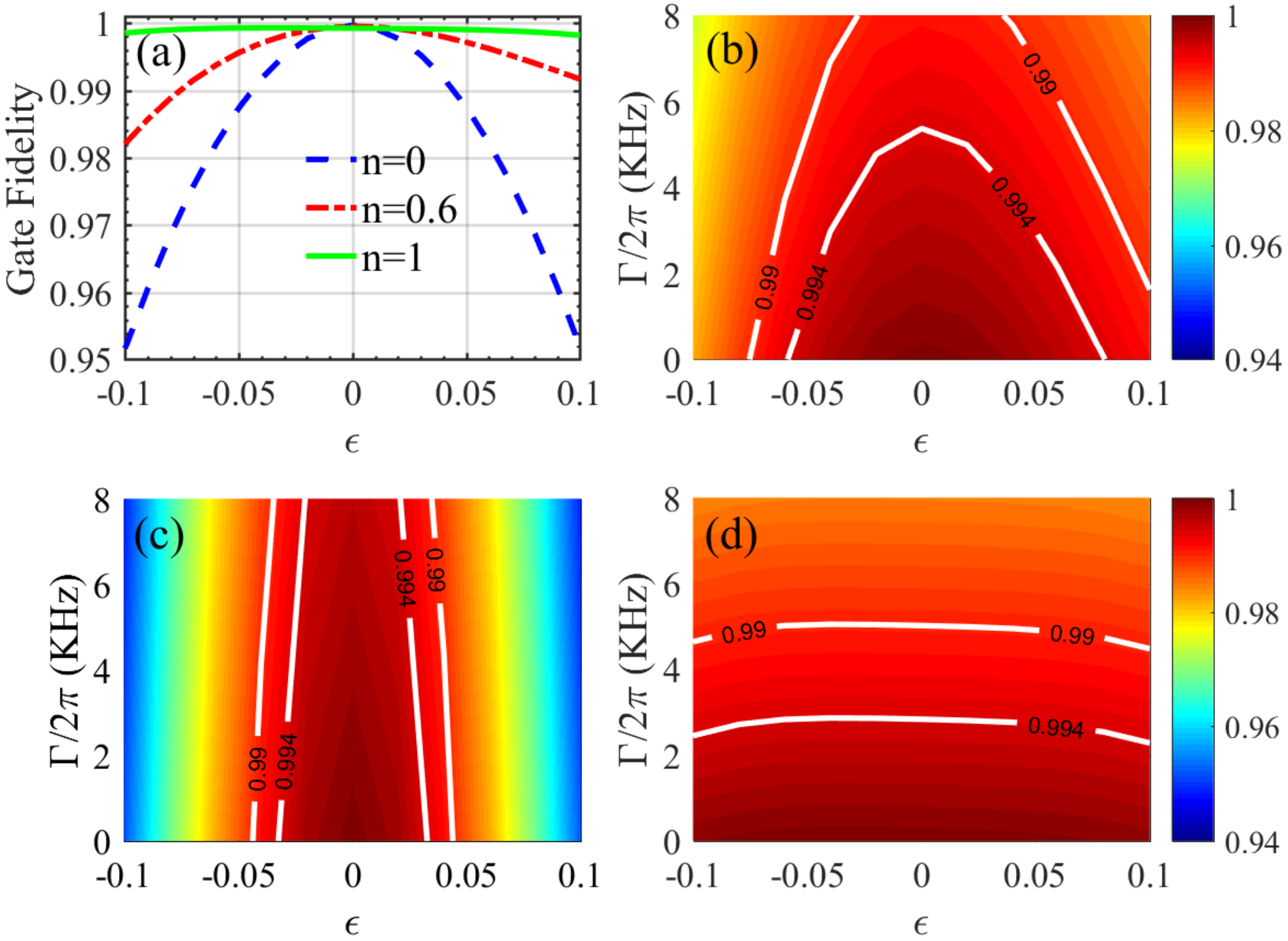}
\caption{Single-qubit gate performance with optimization. (a) The gate fidelity dynamics of the NOT gate with different values n under the systematic error $\epsilon$ without decoherence. (b), (c), and (d) respectively exhibit the gate fidelity dynamics of the NOT gate with optimal value of n=0.6, previous NHQC scheme of n=0 and $q_s = 0$ case of n=1 considering both the systematic error $\epsilon$ and decoherence $\Gamma$.  }
	\end{center}
\end{figure}

However, in the case of  $n \geq 1$, under the restriction, the evolution time $\tau$ will be too long, and decoherence will introduce unacceptable gate infidelity. Therefore, we need to confirm the optimal value of $n$ under the targets with  both short time $\tau$ and low systematic error sensitivity.
To find out the optimal value $n$ under decoherence, we simulated the NOT gate fidelity under the systematic error $-0.1\leq\epsilon\leq0.1$ while changing $n$ from 0 to 1 with the uniform step $dn=0.1$. In this way, we find out that $n=0.6$ is an optimal value. In Fig. 3(a), we plot the gate fidelity in the case $n=1$, $n=0.6$, $n=0$ without  decoherence, respectively. From the Fig. 3(a), we find out that the robustness of the holonomic quantum gates in our scheme is significant improved comparing $n=1$ with $n=0$, corresponding to the previous  NHQC scheme. From Fig. 3(b), (c) and (d), considering both the systematic error and the decoherence effect, we find that optimal value of $n=0.6$ is better. Furthermore, for the long coherence time quantum systems, as shown in Fig. 3(d),  our scheme will significantly improve the robustness of the holonomic quantum gates.

\section{Nontrivial two-qubit gates}
In this section, we proceed to implement nontrivial two-qubit quantum gates. Based on the current experimental technique, the strong capacity coupling between transmon qubits has been achieved experimentally on superconducting circuits  \cite{Capacity0,Capacity1,Capacity2,Capacity3}. Here, we consider the case that  two transmon qubits are capacitively coupled simultaneously to an auxiliary transmon by the capacity coupling. As shown in Fig. 1(c),  the auxiliary transmon with frequency $\omega_A$ dispersively coupled to both qubits with frequencies $\omega_{ge}^k$ $(k=1,2)$.
Meanwhile, the sequential transitions of  both qubits  are driven by microwave field with time dependent driving amplitude $\tilde{\Omega}_k(t)$,   frequency $\tilde{\omega}_k(t)$ and phase $  \tilde{\phi}_k(t)$, i.e. $F_k=\tilde{\Omega}_k(t)\cos[\tilde{\omega}_k(t)+ \tilde{\phi}_k(t)]$. In the rotating framework with respect to the driving frequency, the Hamiltonian of the $k$th  qubit  coupled to the auxiliary transmon can be written as
\begin{eqnarray}\label{Hor}
{H_0} &=& {\delta _k}{N_k} -  \frac{\alpha_k }{2}\left( {{N_k} - 1} \right){N_k} +{\delta _A}{N_A}  -  \frac{\alpha_A }{2}\left( {{N_A} - 1} \right){N_A},\notag\\
H' &=& g_k a{b^\dag_k } + \frac{{\tilde{\Omega}_k {e^{i\tilde{\phi}_k }}}}{2}b_k + \text{H.c.},
\end{eqnarray}
where $H_0$ is the free term of the Hamiltonian for the coupled system, the first two terms represent the nonlinear energy levels of the $k$th  qubit with anharmonicity $\alpha_k$, and the last two terms represent the nonlinear energy levels of the auxiliary transmon with anharmonicity $\alpha_A$, and $\delta_k=\omega_{ge}^k-\tilde{\omega}_k$, $\delta_A=\omega_A-\tilde{\omega}_k$, $N_A = a^{\dagger} a, N_k = b^{\dagger}_kb_k$, with $a=|g\rangle_A\langle e|+\sqrt{2}|e\rangle_A\langle f|+\sqrt{3}|f\rangle_A\langle h|+...$, $b_k=|g\rangle_k\langle e|+\sqrt{2}|e\rangle_k\langle f|+\sqrt{3}|f\rangle_k\langle h|+...$ being the lower operator for the auxiliary transmon and the qubits. $H'$ is the linear interaction term of the system with the coupling strength $g_k$ between the $k$th  qubit and the auxiliary transmon. As all the sequential transitions are allowed in both qubits, the effective interaction will be generated from the interference of the two paths. In addition, the two couplings form a two-photon resonant situation, i.e., $\omega_{ge}^k-\omega_A = \tilde{\omega}_k-\omega_{ef}^k=\Delta>\alpha_k$, and thus lead to driving-assisted coherent resonant coupling between the auxiliary transmon and the $|g\rangle_k\leftrightarrow|f\rangle_k$ transition of the qubits (see Appendix A for details), the Hamiltonian may be expressed as
 \begin{eqnarray}\label{jc}
	{{\tilde H}_2} &=& {\eta_{ge}}\left| {g,e} \right\rangle_{k,A} \left\langle {g,e} \right|
	+ {\eta_{fg}}\left| {f,g} \right\rangle_{k,A} \left\langle {f,g} \right|\notag\\
	&&+ \left( \tilde{g}_k e^{-i\tilde{\phi}_k} \left| {f,g} \right\rangle _{k,A} \left\langle {g,e} \right| + \text{H.c.} \right),
\end{eqnarray}
where
\begin{eqnarray}
{\eta_{ge}}&=& \frac{{{\tilde{\Omega} ^2_k}}}{{4\left( {\Delta-\alpha_k }\right)}}-\frac{{{g^2_k}}}{\Delta },\notag \\
{\eta_{fg}} &=&\frac{{3{\tilde{\Omega} ^2_k}}}{4({\Delta + \alpha_k })} + \frac{{2{g^2_k}}}{{\Delta  - \alpha_k }}
  -\frac{{{\tilde{\Omega} ^2_k}}}{{2\Delta }}, \\
\tilde{g}_k  &=& \frac{{ \sqrt 2 g_k\tilde{\Omega }_k}}{2(\Delta - \alpha_k)}-\frac{{\sqrt 2 g_k \tilde{\Omega}_k}}{2\Delta }
	= \frac{{g_k\tilde{\Omega}_k \alpha_k }}{{\sqrt 2 \Delta \left( {\Delta  - \alpha_k } \right)}},\notag
\end{eqnarray}
where the effective resonant interaction strength $\tilde{g}_k$ is induced by the interference of two Raman-like paths, as shown in Fig. 1(c), similar to Ref. \cite{jc}, and itself and the  phase $\tilde{\phi}_k $ can be tunable via tuning the amplitude and phase of the driving field $F_k$, respectively. Therefore, they can be fine-tuned to meet the requirement for implementing the optimal control in the case of the two-qubits gates.

\begin{figure}[tbp]\label{Figure4}
\begin{center}
\includegraphics[width=6.5cm]{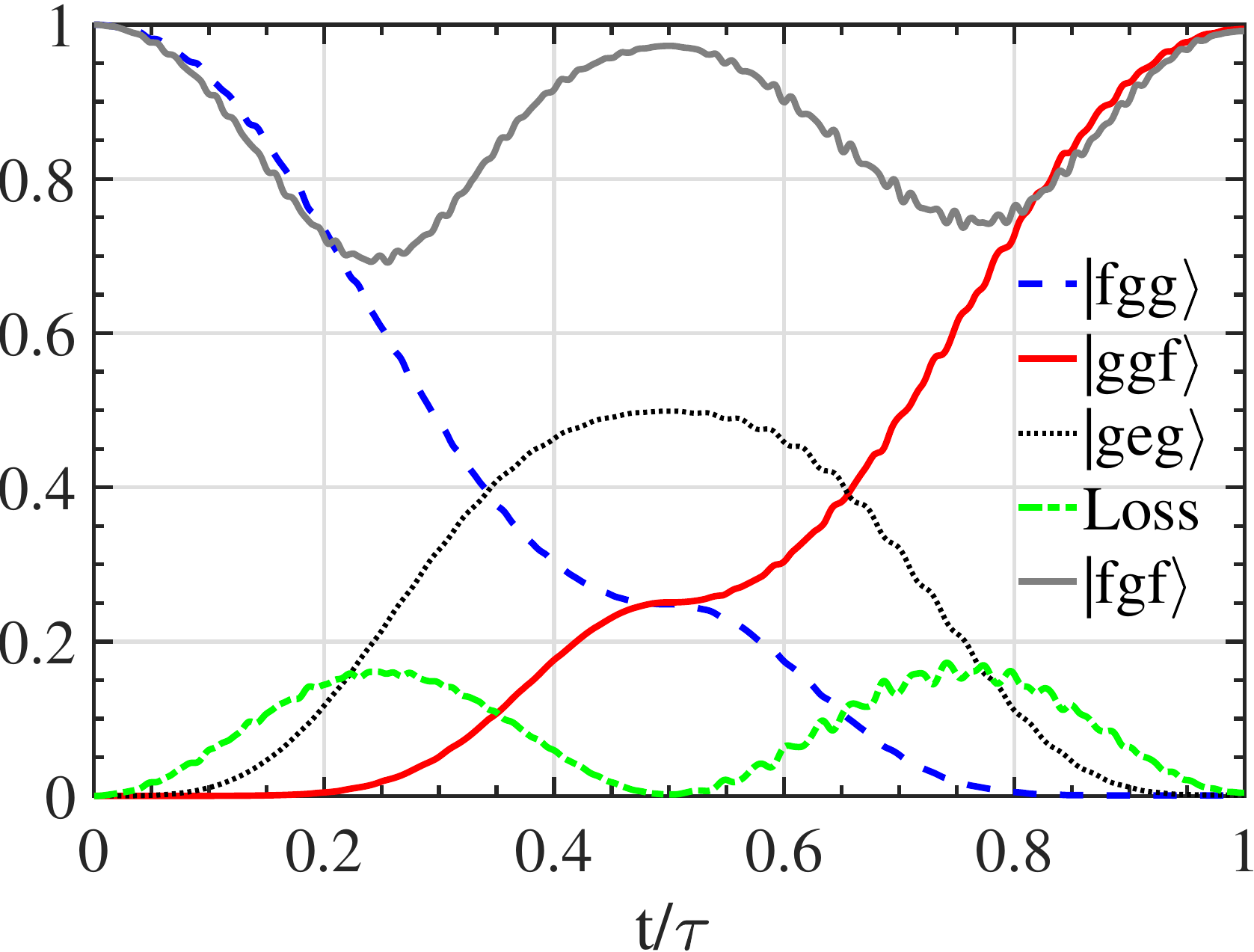}
\caption{Two-qubit gate performance. State population and fidelity dynamics for gate as a function of time with the initial state being $|fgg\rangle$ with Loss and initial state being $ |fgf\rangle$.}
\end{center}
\end{figure}

We set $\Delta \gg \{g_k, \tilde{\Omega}_k\}$, after concealing  the cross-ac-Stark-shifts by modulating the frequencies of the driving fields accordingly to $\tilde{\Omega}_k$ (see Appendix B for details), in the single-excitation subspace $S_1=$span$\{|ggf\rangle,|fgg\rangle,|geg\rangle\}$, where $|lms\rangle \equiv |l\rangle_1\otimes|m\rangle_A\otimes|s\rangle_2$ labels the product states of the two qubits and the auxiliary transmon,  the effective interaction Hamiltonian can be described by
\begin{eqnarray}\label{coupled}
H_{eff} &=& \tilde{g}_1e^{-i\tilde{\phi}_1}  \left| {fgg} \right\rangle \left\langle {geg} \right| + \tilde{g}_2e^{-i\tilde{\phi}_2}  \left| {ggf} \right\rangle \left\langle {geg} \right|+ \mathrm{H.c.} \notag\\
&=& \tilde{g}e^{-i{\tilde{\phi}}_1}|B\rangle\langle E| + \mathrm{H.c.},
\end{eqnarray}
where $\tilde{g} =\sqrt{\tilde{g}^2_1+\tilde{g}^2_2}$, bright state $|B\rangle=\sin\frac{\vartheta}{2}|fgg\rangle -e^{-i\tilde{\phi}}\cos\frac{\vartheta}{2}|ggf\rangle$ with $\tan(\vartheta/2)=\tilde{g}_1/\tilde{g}_2$ and $\tilde{\phi} = \tilde{\phi}_2-\tilde{\phi}_1+\pi$,  $|E\rangle=|geg\rangle$, and dark state $|D\rangle=-\cos(\vartheta/2)e^{i\tilde{\phi}}|fgg\rangle -\sin(\vartheta/2)|ggf\rangle$ is decoupled. The above effective Hamiltonian, which can readily be used to implement nontrivial two-qubit gates, establishes a equivalent three-level Hamiltonian in the single-excitation subspace with  $|E\rangle$ being an auxiliary state,  as illustrated in Fig. 1(d). Then, we can adopt the same protocol as for the single-qubits case to implement holonomic two-qubit gates. Notably, $\tilde{g}$ and ${\tilde{\phi}_1}$ in the effective Hamiltonian in Eq.  (\ref{coupled}) can also be solved by Eq. (\ref{SE}) similar to the single-qubits case, i.e.
\begin{equation}\label{Jp2}
\tilde{\phi}_1 =\varphi^\prime-  \arctan{\left(\frac{\dot{\chi}^\prime}{\dot{\varphi^\prime}}\cot{\chi^\prime}\right)},
\tilde{g} = \frac{\dot{\chi^\prime}}{2\sin{(\tilde{\phi}_1-\varphi^\prime)}},
\end{equation}
 where $\chi^\prime$ and $\varphi^\prime$ are chosen the same form in Eq. (\ref{parameter1}) and Eq. (\ref{parameter2}), after that, the effective Hamiltonian can be fixed. Thus, for the case of $\tilde{\gamma}=\pi,\tilde{\phi}=0$, the evolution operator in the two-qubit gate Hilbert space $S_2=$span$\{|gg\rangle,|gf\rangle,|fg\rangle,|ff\rangle\}$ can be written as \begin{eqnarray}\label{UUU}
 U_2(\vartheta) &=& \left(\begin{array}{cccc}
 1 &0 &0 &0 \\
 0 &\cos{\vartheta}&\sin{\vartheta}&0\\
 0 &\sin{\vartheta} &-\cos{\vartheta}&0\\
 0 &0 &0&1\\
 \end{array}\right).
 \end{eqnarray}
Now, we analyse the performance of two-qubit gates with $\vartheta=\pi/2$. For $\Delta = 2\pi \times 1 $ GHz, and the parameter of transmons $g_k = 2\pi \times 65$ MHz, $\alpha_k = 2\pi \times 400$ MHz, $\alpha_A = 2\pi \times 370$ MHz,  $\tau_2 = 57$ ns, $\tilde{g}_{k, max} = 10$ MHz by modulating $\tilde{\Omega}_k(t)$ with the maximum value to be $2\pi \times 320$ MHz. When the initial state  is $|fgg\rangle$, a fidelity of $99.44\%$ can be obtained, as plotted in Fig. 4, which is done by using the origin Hamiltonian in Eq.(\ref{Hor}), i.e., including all the unwanted higher-order effects induced by the strong microwave driving.  In addition, the loss represents the leakage from our computational basis to neighboring states caused directly by the time dependence of the amplitude of the driving pulse, leading to the time dependence of the cross-ac-Stark-shifts terms, which can be compensated by modulating the pulse frequencies $\tilde{\omega}_k$ accordingly. Notably,  due to the amplitude of the driving pulse being zero at the beginning and the ending of the evolution, as we conceal the cross-ac-Stark-shifts, the loss is zero before and after the operation. Similarly, the optimal control technique presented in the single-qubit implementation can also be incorporated in this  two-qubit gate implementation.

\section{Discussion and conclusion}
In conclusion, we have proposed a practical  implementation of fast NHQC on capacitively coupled superconducting circuits with microwave fields induced effective resonant coupling. Our scheme can be scaled up to two-dimensional square lattice of qubits, which is a scalable setup for QC. Besides the scalability, our scheme has the following additional distinct merits. First, instead of linear resonators, auxiliary transmons are used to induced nontrivial two-qubit gates, the intrinsic nonlinearity of which can be large enough for our purpose. On the other hand, using the transmons to serve as auxiliary elements makes our scalable setup consists only the qubits, and thus easier to be fabricated. Second, due to the fine inverse Hamiltonian engineering, the evolution state of our proposal has two independent variables, thus it meets the minimal requirement of optimal control techniques, which can be used to further enhances the robustness of our implementation. Therefore, our scheme removes the main obstacles of NHQC, and thus provides a promising way towards robust NHQC on superconducting circuits.

\appendix
\section*{Appendix}
\section{The effective Hamiltonian}
Starting from the original Hamiltonian of Eq. (\ref{Hor}) in the main text, the energies of the state $\left| {g,e} \right\rangle ,\left| {f,g} \right\rangle $ are
\begin{equation}
	{E_{f,g}} = 2{\delta _k} - \alpha_k,  \quad
	{E_{g,e}} = {\delta _A},
\end{equation}
which can be adjusted to be degenerate by modulating $\tilde{\omega}_k$ such that ${\delta _A} = 2{\delta _k} - \alpha_k$, and set $\varepsilon =E_{f,g}=E_{g,e}$.
We then define
\begin{eqnarray}
	\mathcal{P} &=& \left| {g,e} \right\rangle_{k,A} \left\langle {g,e} \right| + \left| {f,g} \right\rangle_{k,A} \left\langle {f,g} \right|,   \\
	\mathcal{K}&=& \sum\limits_\Pi  {\frac{{\left| {l,m} \right\rangle_{k,A} \left\langle {l,m} \right|}}{{{\varepsilon _{l,m}} - \varepsilon }}},
\end{eqnarray}
where the degenerate subspace $\{\left| {g,e} \right\rangle_{k,A}, \left| {f,g} \right\rangle_{k,A}   \}$ is of interest, $\Pi :\left\{ {\left. {l,m} \right|\left( {l,m} \right) \ne \left( {g,e} \right)or\left( {f,g} \right)} \right\}$, and $\varepsilon _{l,m}$ is the energy of state $\left| {l,m} \right\rangle$. In the following, we only consider the fourth energy level that is beyond the qubit states.

We handle the effective Hamiltonian using a perturbation theory with $\{g_k, \tilde{\Omega}_k\} \ll \Delta=\delta_k-\delta_A$, the first-order term is found to be
\begin{equation}
	{{\tilde H}_1} = \mathcal{P}H'\mathcal{P}=0,
\end{equation}
as
\begin{eqnarray}
H'\mathcal{P} 	&= &\left( { g_k\left| {e,g} \right\rangle  + \frac{\tilde{\Omega}_k }{2}{e^{ - i\tilde{\phi}_k }}\left| {e,e} \right\rangle } \right)_{k,A}\left\langle {g,e} \right|  \notag\\
& +&\left(\sqrt 2 g_k\left| {e,e} \right\rangle  + \frac{{\sqrt 2  \tilde{\Omega}_k }}{2}{e^{i\tilde{\phi}_k }}\left| {e,g} \right\rangle \right. \notag\\
&&+ \left.\frac{{\sqrt 3  \tilde{\Omega}_k {e^{ - i\tilde{\phi}_k }}}}{2}\left| {h,g} \right\rangle  \right)_{k,A}\left\langle {f,g} \right|.
\end{eqnarray}
As for the second-order terms,
\begin{eqnarray}
	{{\tilde H}_2} &=& - \mathcal{P}H'\mathcal{K}H'\mathcal{P} \notag \\
	&=& - \mathcal{P}H'\left( \mathcal{K}_1 + \mathcal{K}_2 + \mathcal{K}_3 \right)H'\mathcal{P},
\end{eqnarray}
where
\begin{eqnarray}
	\mathcal{K}_1 &=& \frac{{\left| {e,g} \right\rangle_{k,A} \left\langle {e,g} \right|}}{{{\varepsilon _{e,g}} - \varepsilon }}, \notag \\
	\mathcal{K}_2 &=& \frac{{\left| {e,e} \right\rangle_{k,A} \left\langle {e,e} \right|}}{{{\varepsilon _{e,e}} - \varepsilon }}, \notag \\
	\mathcal{K}_3 &=& \frac{{\left| {h,g} \right\rangle_{k,A} \left\langle {h,g} \right|}}{{{\varepsilon _{h,g}} - \varepsilon }}. \notag
\end{eqnarray}
Finally, we get the Eq.(\ref{jc}) in main text.
\begin{figure}[tbp]\label{Figure5}
\includegraphics[width=6.5cm]{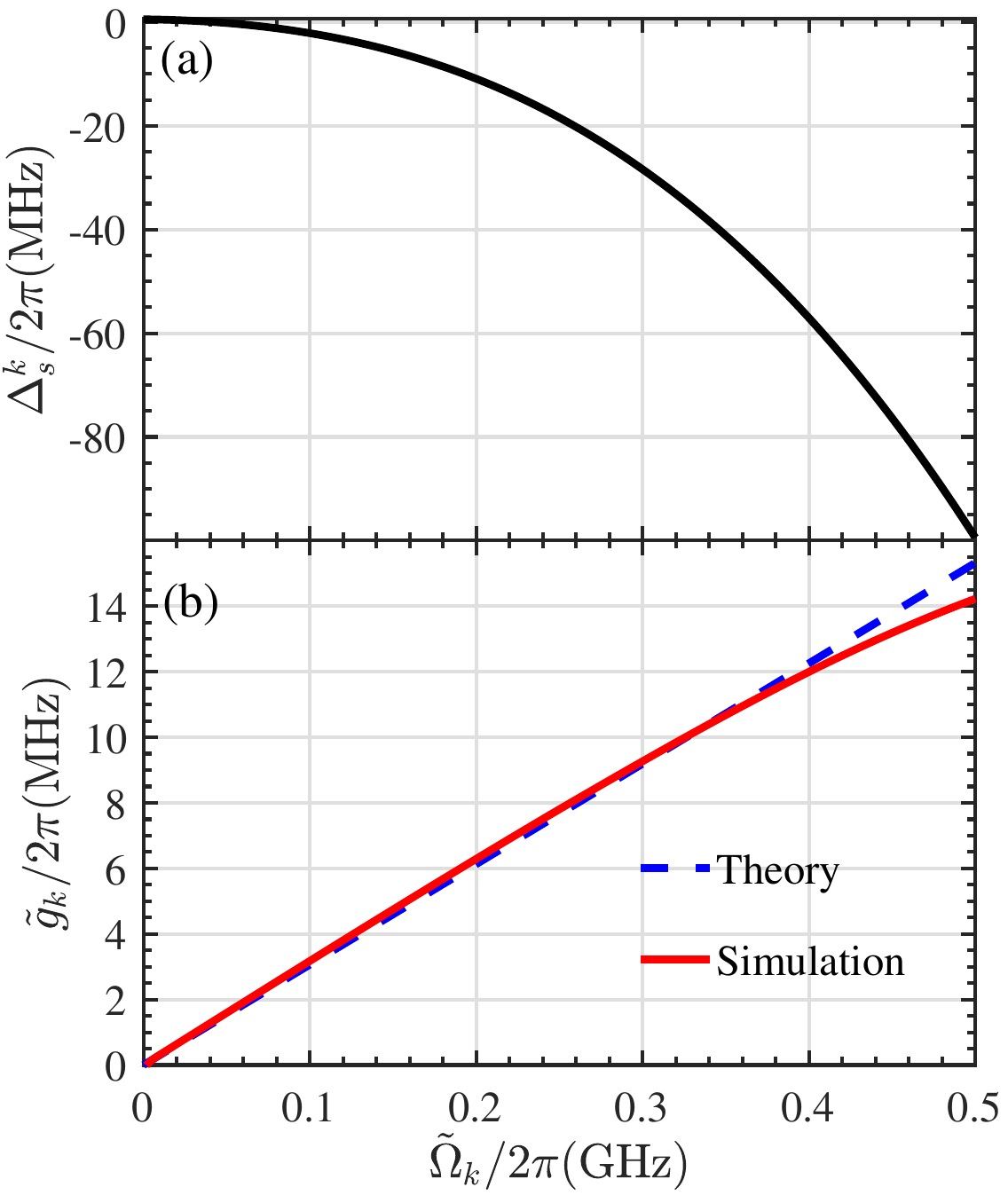}
\caption{(a) Illustration of the $\Delta^k_s$  with respect to $\tilde{\Omega}_k$ for
fixed $g_k$. (b) Illustration of the effective transmon-transmon coupling strength $\tilde{g}$ with respect to $\tilde{\Omega}_k$ with fixed $g_k$.}
\end{figure}

\section{Compensate of the cross-ac-Stark-shifts}
In Eq. (\ref{jc}), there are just ac Stark shifts caused by the $k$th  qubit coupled to an auxiliary transmon. When we consider the two qubits simultaneously coupled to an auxiliary transmon, the cross-ac-Stark-shifts will occur \cite{dje2019}. In the degenerate subspace $S_1$, the whole Hamiltonian can be written as
\begin{eqnarray}\label{Hcross}
{H}_{\mathrm{two}}&=& \eta_{fgg} | fgg \rangle \langle fgg | +\eta_{{geg}}|geg\rangle\langle geg| \notag\\
&&+\eta_{{ggf}}|ggf\rangle\langle ggf| + \tilde{g}_{1}e^{-i\tilde{\phi}_1} | fgg \rangle\langle geg |\notag\\
&&+\left(\tilde{g}_{2}e^{-i\tilde{\phi}_2}| ggf\rangle\langle geg |+\mathrm{H.c.}\right) \notag\\
&=& \eta_{B} | B \rangle \langle B |+\eta_{D} | D \rangle \langle D | +\eta_{{E}}|E\rangle\langle E|\notag\\
&&+ \left(\tilde{g}e^{-i\tilde{\phi}_1} | B \rangle\langle E |+\mathrm{H.c.}\right),
\end{eqnarray}
where $\eta_{{M}}$ is the energy level shift of the state $| {M} \rangle$. Due to the existence of the cross-ac-Stark-shifts, it can lead to large errors of the gate operations. Therefore, we need to compensate these shifts. It is noted that both $g_k$ and $\tilde{\Omega}_k$ split the degenerate subspace $\{| B \rangle, | D \rangle, | E \rangle\}$, so we will fix $g_k$ and tune the frequency $\tilde{\omega}_k$ of the driven field to fulfill $|\eta_{B}-\eta_{E}|=|\eta_{B}-\eta_{D}|=0$. However, as $|\eta_{B}-\eta_{E}|\geq|\eta_{B}-\eta_{D}| $ due to $\tilde{g}$, we just need to care $|\eta_{B}-\eta_{E}|=0$ in the degenerate subspace $\{| B \rangle, | E \rangle\}$,  i.e.,
$$\left\langle {{\phi _l}\left( \tilde{\Omega}_k  \right)} \right|\frac{d}{{d\tilde{\Omega}_k \left( t \right)}}\left| {{\phi _m}\left(\tilde{\Omega}_k  \right)} \right\rangle  = 0.$$
Then, we obtain
\begin{eqnarray}
&&\left\langle {{\phi _l}} (\tilde{\Omega}_k )\right|\frac{{\partial H}}{{\partial \tilde{\Omega}_k }}\left| {{\phi _m}} (\tilde{\Omega}_k )\right\rangle \notag\\
&+& \frac{{d\tilde{\omega}_k }}{{d\tilde{\Omega}_k }}\left\langle {{\phi _l}} (\tilde{\Omega}_k ) \right|\frac{{\partial H}}{{\partial \tilde{\Omega}_k }}\left| {{\phi _m}} (\tilde{\Omega}_k ) \right\rangle  = 0,
\end{eqnarray}
which can be numerically solved to obtain the $\tilde{\omega}_k-\tilde{\Omega}_k$ curve, such that one can figure out the $\Delta^k_s = \tilde{\omega}_k -\tilde{\omega}_k(0)$ under the situation $\tilde{\Omega}_1=\tilde{\Omega}_2 $ for numerical simulation, as shown in Fig. 5(a). Once the above equation is satisfied, the cross-ac-Stark-shifts  must be compensated by tuning the frequency $\tilde{\omega}_k$ of the driven field.

After that, in order to effectively conceal the cross-ac-Stark-shifts, the driven pulse with smoothly changed amplitude can be employed, which can lead to a smoothly changed effectively resonant coupling strength $\tilde{g}_k$. As shown in Fig. 5(b), the $\tilde{g}_k-\tilde{\Omega}_k$ curve of the numerical simulation is coincident to the theory. Also note that when $\tilde{\Omega}_k$ is large, the trend of $\tilde{g}_k-\tilde{\Omega}_k$ will be slightly nonlinear.

\section*{Acknowledgments}

This work was supported by the Key-Area Research and Development Program of GuangDong Province (Grant No. 2018B030326001), the National Natural Science Foundation of China (Grant No. 11874156), the National Key R\&D Program of China (Grant No. 2016 YFA0301803), and the  research project from SCNU (Grant No. 19WDGB04).

\section*{Conflict of Interest}
The authors declare no conflict of interest.


\begin{thebibliography}{00}

\bibitem{berry} M. V. Berry, \textit{Proc. R. Soc. Lond., Ser. A} \textbf{1984}, \textit{392}, 45.

\bibitem{b3} F. Wilczek, A. Zee, \textit{Phys. Rev. Lett.} \textbf{1984},  \textit{52}, 2111.

\bibitem{b2} Y. Aharonov, J. Anandan, \textit{Phys. Rev. Lett.} \textbf{1987}, \textit{58}, 1593.

\bibitem{ps1} P. Solinas, P. Zanardi, N. Zangh\`{\i}, \textit{Phys. Rev. A} \textbf{2004}, \textit{70}, 042316.

\bibitem{zhu05} S. L. Zhu, P. Zanardi, \textit{Phys. Rev. A} \textbf{2005}, \textit{72}, 020301(R).

\bibitem{ps2} P. Solinas, M. Sassetti, T. Truini, N. Zangh\`{\i}, \textit{New J. Phys.} \textbf{2012}, \textit{14}, 093006.

\bibitem{mj} M. Johansson, E. Sj\"{o}qvist, L. M. Andersson, M. Ericsson, B. Hessmo, K. Singh, D. M. Tong, \textit{Phys. Rev. A} \textbf{2012}, \textit{86}, 062322.

\bibitem{gqc} E. Sj\"{o}qvist, \textit{Physics} \textbf{2008}, \textit{1}, 35.

\bibitem{zanardi} P. Zanardi, M. Rasetti, \textit{Phys. Lett. A} \textbf{1999}, \textit{264}, 94.

\bibitem{xbwang} X. B. Wang, M. Keiji, \textit{Phys. Rev. Lett.} \textbf{2001}, \textit{87}, 097901.

\bibitem{zhu} S. L. Zhu, Z. D. Wang, \textit{Phys. Rev. Lett.}  \textbf{2002}, \textit{89}, 097902.

\bibitem{Sjoqvist2012} E. Sj\"{o}qvist, D. M. Tong, L. M. Andersson, B. Hessmo, M. Johansson, K. Singh, \textit{New J. Phys.} \textbf{2012}, \textit{14}, 103035.

\bibitem{Xu2012} G. F. Xu, J. Zhang, D. M. Tong, E. Sj\"{o}qvist, L. C. Kwek, \textit{Phys. Rev. Lett.} \textbf{2012}, \textit{109}, 170501.

\bibitem{vam2014} V. A. Mousolou, E. Sj\"{o}qvist, \textit{Phys. Rev. A} \textbf{2014}, \textit{89}, 022117.

\bibitem{Zhang2014d} J. Zhang, L. C. Kwek, E. Sj\"{o}qvist, D. M. Tong, P. Zanardi, \textit{Phys. Rev. A} \textbf{2014}, \textit{89}, 042302.

\bibitem{Xu2014} G. F. Xu, G. L. Long, \textit{Sci. Rep.} \textbf{2014}, \textit{4}, 6814.

\bibitem{Xu2015} G. F. Xu, C. L. Liu, P. Z. Zhao, D. M. Tong, \textit{Phys. Rev. A}  \textbf{2015},  \textit{92}, 052302.

\bibitem{Xue2015b} Z. Y. Xue, J. Zhou, Z. D. Wang, \textit{Phys. Rev. A} \textbf{2015}, \textit{92}, 022320.

\bibitem{es2016} E. Sj\"{o}qvist, \textit{Phys. Lett. A} \textbf{2016}, \textit{380}, 65.

\bibitem{Xue2016} Z. Y. Xue, J. Zhou, Y. M. Chu, Y. Hu, \textit{Phys. Rev. A} \textbf{2016}, \textit{94}, 022331.

\bibitem{Zhao2016} P. Z. Zhao, G. F. Xu, D. M. Tong, \textit{Phys. Rev. A} \textbf{2016}, \textit{94}, 062327.

\bibitem{Herterich2016} E. Herterich, E. Sj\"{o}qvist, \textit{Phys. Rev. A} \textbf{2016}, \textit{94}, 052310.

\bibitem{xu2017} G. F. Xu, P. Z. Zhao, T. H. Xing, E. Sj\"{o}qvist, D. M. Tong, \textit{Phys. Rev. A} \textbf{2017}, \textit{95}, 032311.

\bibitem{Xue2017} Z. Y. Xue, F. L. Gu, Z. P. Hong, Z. H. Yang, D. W. Zhang, Y. Hu, J. Q. You, \textit{Phys. Rev. Appl.} \textbf{2017}, \textit{7}, 054022.

\bibitem{xu20172} G. F. Xu, P. Z. Zhao, D. M. Tong, E. Sj\"{o}qvist, \textit{Phys. Rev. A} \textbf{2017}, \textit{95}, 052349.

\bibitem{zhao2017} P. Z. Zhao, G. F. Xu, Q. M. Ding, E. Sj\"{o}qvist, D. M. Tong, \textit{Phys. Rev. A} \textbf{2017}, \textit{95}, 062310.

\bibitem{vam2017} V. A. Mousolou, Phys. Rev. A \textbf{2017}, \textit{96}, 012307.

\bibitem{zhouj2018} J. Zhou, B. J. Liu, Z. P. Hong, Z. Y. Xue, \textit{Sci. China: Phys. Mech. Astron.} \textbf{2018}, \textit{61}, 010312.

\bibitem{Hong18} Z. P. Hong, B. J. Liu, J. Q. Cai, X. D. Zhang, Y. Hu, Z. D. Wang, Z. Y. Xue, \textit{Phys. Rev. A} \textbf{2018}, \textit{97}, 022332.

\bibitem{xu2018} G. F. Xu, D. M. Tong, E. Sj\"{o}qvist, \textit{Phys. Rev. A} \textbf{2018}, \textit{98}, 052315.

\bibitem{vam2018} V. A. Mousolou, \textit{Phys. Rev. A} \textbf{2018}, \textit{98}, 062340.

\bibitem{Zhang2019} F. Zhang, J. Zhang, P. Gao, G. Long, \textit{Phys. Rev. A} \textbf{2019}, \textit{100}, 012329.

\bibitem{nr2019} N. Ramberg, E. Sj\"{o}qvist, \textit{Phys. Rev. Lett.} \textbf{2019}, \textit{122}, 140501.

\bibitem{Abdumalikov2013} A. A. Abdumalikov, J. M. Fink, K. Juliusson, M. Pechal, S. Berger, A. Wallraff, S. Filipp, \textit{Nature (London)} \textbf{2013}, \textit{496}, 482.

\bibitem{Feng2013} G. Feng, G. Xu, G. Long, \textit{Phys. Rev. Lett.} \textbf{2013}, \textit{110}, 190501.

\bibitem{Zu2014} C. Zu, W. B. Wang, L. He, W. G. Zhang, C. Y. Dai, F. Wang, L. M. Duan, \textit{Nature (London)} \textbf{2014}, \textit{514}, 72.

\bibitem{Arroyo-Camejo2014} S. Arroyo-Camejo, A. Lazariev, S. W. Hell, G. Balasubramanian, \textit{Nat. Commun.} \textbf{2014}, \textit{5}, 4870.


\bibitem{nv2017} Y. Sekiguchi, N. Niikura, R. Kuroiwa, H. Kano, H. Kosaka, \textit{Nat. Photonics}  \textbf{2017}, \textit{11}, 309.

\bibitem{nv20172} B. B. Zhou, P. C. Jerger, V. O. Shkolnikov, F. J. Heremans, G. Burkard, D. D. Awschalom, \textit{Phys. Rev. Lett.} \textbf{2017}, \textit{119}, 140503.


\bibitem{li2017}  H. Li, L. Yang, G. Long, \textit{Sci. China: Phys., Mech. Astron.}  \textbf{2017}, \textit{60}, 080311.

\bibitem{Xu18}
Y. Xu, W. Cai, Y. Ma, X. Mu, L. Hu, T. Chen, H. Wang, Y. P. Song, Z. Y. Xue, Z. Q. Yin, L. Sun, \textit{Phys. Rev. Lett.} \textbf{2018}, \textit{121}, 110501.

\bibitem{ni2018}
N. Ishida, T. Nakamura, T. Tanaka, S. Mishima, H. Kano, R. Kuroiwa, Y. Sekiguchi, H. Kosaka, \textit{Opt. Lett.} \textbf{2018}, \textit{43}, 2380.

\bibitem{kn2018} K. Nagata, K. Kuramitani, Y. Sekiguchi, H. Kosaka, \textit{Nat. Commun.}  \textbf{2018}, \textit{9}, 3227.

\bibitem{dje2019} D. J. Egger, M. Ganzhorn, G. Salis, A. Fuhrer, P. Muller, P. K. Barkoutsos, N. Moll, I. Tavernelli, S. Filipp, \textit{Phys. Rev. Appl.} \textbf{2019}, \textit{11}, 014017.

\bibitem{yan2019} T. Yan, B. J. Liu, K. Xu, C. Song, S. Liu, Z. Zhang, H. Deng, Z. Yan, H. Rong, K. Huang, M. H. Yung, Y. Chen, D. Yu, \textit{Phys. Rev. Lett.}  \textbf{2019}, \textit{122}, 080501.

\bibitem{zhu2019} Z. Zhu, T. Chen, X. Yang, J. Bian, Z. Y. Xue, X. Peng, \textit{Phys. Rev. Appl.}  \textbf{2019}, \textit{12}, 024024.


\bibitem{sq2} J. Clarke, F. K. Wilhelm, \textit{Nature (London)} \textbf{2008}, \textit{453}, 1031.

\bibitem{sq3} J. Q. You, F. Nori, \textit{Nature (London)} \textbf{2011}, \textit{474}, 589.

\bibitem{sq4} M. H. Devoret, R. J. Schoelkopf, \textit{Science} \textbf{2013}, \textit{339}, 1169.

\bibitem{transmon} J.  Koch,  T. M. Yu, J. Gambetta, A. A. Houck, D. I. Schuster, J. Majer, A. Blais, M. H. Devoret, S. M. Girvin, R. J. Schoelkopf, \textit{Phys. Rev. A} \textbf{2007}, \textit{76}, 042319.

\bibitem{multilevel} M. J. Peterer, S. J. Bader, X. Jin, F. Yan, A. Kamal, T. J. Gudmundsen, P. J. Leek, T. P. Orlando, W. D. Oliver, S. Gustavsson, \textit{Phys. Rev. Lett.} \textbf{2015}, \textit{114}, 010501.


\bibitem{Zheng16} S. B. Zheng, C. P. Yang, F. Nori, \textit{Phys. Rev. A} \textbf{2016}, \textit{93}, 032313.

\bibitem{Jing17}  J. Jing, C. H. Lam, L. A. Wu, \textit{Phys. Rev. A} \textbf{2017}, \textit{95}, 012334.

\bibitem{Liu18} B. J. Liu, X. K. Song, Z. Y. Xue, X. Wang, M. H. Yung, \textit{Phys. Rev. Lett.} \textbf{2019}, \textit{123}, 100501.

\bibitem{Daems13} D. Daems, A. Ruschhaupt, D. Sugny, S. Gu\'{e}rin, \textit{Phys. Rev. Lett.}  \textbf{2013}, \textit{111}, 050404.


\bibitem{Chen12} A. Ruschhaupt, X. Chen, D. Alonso, J. G. Muga, \textit{New J. Phys.} \textbf{2012}, \textit{14}, 093040.


\bibitem{Barends14} R. Barends, J. Kelly, A. Megrant, A. Veitia, D. Sank, E. Jeffrey, T. C. White, J. Mutus, A. G. Fowler, B. Campbell, Y. Chen, Z. Chen, B. Chiaro, A. Dunsworth, C. Neill, P. O¡¯Malley, P. Roushan, A. Vainsencher, J. Wenner, A. N. Korotkov, A. N. Cleland, J. M. Martinis, \textit{Nature (London)} \textbf{2014}, \textit{508}, 500.



\bibitem{jc} S. Zeytino\u{g}lu, M. Pechal, S. Berger, A. A. Abdumalikov Jr., A. Wallraff, S. Filipp, \textit{Phys. Rev. A} \textbf{2015}, \textit{91}, 043846.

\bibitem{Poyatos97} J. F. Poyatos, J. I. Cirac, P. Zoller, \textit{Phys. Rev. Lett.} \textbf{1997}, \textit{78}, 390.

\bibitem{Capacity0} M. H. Goerz, F. Motzoi, K. B. Whaley, C. P. Koch, \textit{npj Quantum Inf.}  \textbf{2017}, \textit{3}, 37.

\bibitem{Capacity1} M. Reagor \emph{et al}., 
    \textit{Sci. Adv.}  \textbf{2018}, \textit{4}, eaao3603.

\bibitem{Capacity2} S. A. Caldwell \emph{et al}., 
    \textit{Phys. Rev. Appl.} \textbf{2018}, \textit{10}, 034050.

\bibitem{Capacity3} X. Li, Y. Ma, J. Han, T. Chen, Y. Xu, W. Cai, H. Wang, Y. P. Song, Z. Y. Xue, Z. Q. Yin, L. Sun, \textit{Phys. Rev. Appl.} \textbf{2018}, \textit{10}, 054009.


\end{thebibliography}
\end{document}